\documentstyle[preprint,aps]{revtex}

\begin{document}

\hsize = 6.5in
\widetext
\draft
\tighten
\topmargin-48pt

\preprint{EFUAZ FT-97-41}

\title{On the way to understanding the electromagnetic phenomena}

\author{{\bf Valeri V. Dvoeglazov}}

\address{
Escuela de F\'{\i}sica, Universidad Aut\'onoma de Zacatecas \\
Apartado Postal C-580, Zacatecas 98068, Zac., M\'exico\\
Internet address:  VALERI@CANTERA.REDUAZ.MX\\
URL: http://cantera.reduaz.mx/\~\,valeri/valeri.htm
}

\author{{\bf Myron W. Evans}}

\address{
Department of Physics and Astronomy, York University\\
4700 Keele Street, Toronto, Ontario M3J 1P3 Canada\\
Internet address: 100561.607@CompuServe.Com\\
URL: http://www.europa.com/\~ \,rsc/physics/B3/Evans
}

\date{January 11, 1997}

\maketitle

\begin{abstract}
On the basis of the ordinary mathematical methods
we discuss new classes of solutions of the
Maxwell's equations discovered  in the papers by D. Ahluwalia, M. Evans
and H. M\'unera {\it et al.}
\end{abstract}


\newpage

\baselineskip13pt

Recently several authors found  additional solutions
to relativistic wave equations. Here they are listed:

\begin{itemize}

\item
$E=0$ solution of the Maxwell's $j=1$ equations~\cite{DVA} which
was found on the basis of the consideration of the characteristic equation
(in the momentum representation).

\item
${\bf B}^{(3)}$ Evans-Vigier field~\cite{EVA}, which was obtained as a
cross-product of the transverse modes of electromagnetism: ${\bf B}^{(1)}
\times {\bf B}^{(2)} = i B^{(0)} {\bf B}^{(3)\, \ast}$ and cyclic.

\item
Non plane-wave solutions of the Klein-Gordon
equation~[3a,b] by M\'unera {\it et al.},
which were obtained by using unconventional basis
functions and ``coupling {\it anzatz}", see~[3a, Eqs. (11,12)].

\item
M\'unera and Guzm\'an generalized solution of Maxwell's equations in terms
of potentials~[3c,d].

\item
Chubykalo and Smirnov-Rueda `method of separated potentials',
ref.~\cite{CHU}, which permits us to consider a function with implicit
dependence on time as full-value solution of the Maxwell's
(and/or D'Alembert) equations.

\end{itemize}

Why did so many new unexpected solutions appear at once?
Let us look at this issue by using ordinary methods of solving the system
of partial differential equations~\cite{TEXTBOOK1,TEXTBOOK2}.

It is well known that the set of dynamical Maxwell's equations are
equivalent to the following set,
{\it e.g.},~\cite[Eqs.(4.21,4.22)]{WEIN}:\footnote{Issues related with
the source equations will be discussed in detail elsewhere.}
\begin{mathletters}
\begin{eqnarray}
\bbox{\nabla} \times [ {\bf E} + i {\bf B} ] - i (\partial / \partial t)
[ {\bf E} + i{\bf B} ] &=&0\quad,\label{1}\\
\bbox{\nabla} \times [ {\bf E} - i {\bf B} ] + i (\partial / \partial t)
[ {\bf E} - i{\bf B} ] &=&0\quad. \label{2}
\end{eqnarray}
\end{mathletters}
This is a system of partial differential equations.
It is easy to see that the second equation is just the parity
conjugate (${\bf x} \rightarrow -{\bf x}$) of the first one if
one uses ordinary interpretation of ${\bf E}$, a {\it vector},
and ${\bf B}$, an {\it axial vector}.

In the framework of this paper we shall look for  solutions of (\ref{1})
in the generalized form\footnote{More rigorous consideration will be
reported in the extended version.} $${\bf A} \equiv {\bf E}+i {\bf B} \sim
{\bf a} \exp (\lambda t + \bbox{\kappa} \cdot {\bf r})\quad,$$ where
$\lambda$ and $\bbox{\kappa}$ are some unknown parameters, which provide
characteristic polynomio, and ${\bf a} = \mbox{column} (a_1\quad a_2 \quad
a_3)$ is some {\it constant} vector, which is defined by the boundary
and/or normalization conditions.  Thus, at the moment we are not going to
restrict our consideration by the plane waves.  As a result of the use of
the method of characteristic polynomio for the differential equation
\begin{equation}
[(\partial /\partial t) + {\bf J}\cdot \bbox{\nabla} ]^{ij} A^j = 0 \quad,
\end{equation} with $({\bf J}^i)^{jk} = -i\epsilon^{ijk}$, we obtain the
algebraic equation for parameters $\lambda$ and $\bbox{\kappa}$:
\begin{equation}
Det [ \lambda
+ ({\bf J} \cdot \bbox{\kappa}) ]^{ij} = 0 \quad.
\end{equation}
It has solutions $\lambda=0$ and $\lambda = \pm \mid \bbox{\kappa} \mid$.
In fact, we repeated the procedure of ref.~\cite{DVA},  but
standing at the most general position we do not know yet, how $\lambda$
and $\bbox{\kappa}$ are connected with energy and momentum. Thus, the
general solution of the first Maxwell equation (\ref{1}) may be presented,
for instance, in the form:
\begin{equation} {\bf E} +i {\bf B} = {\bf
A}_1 \exp [\alpha_1 (\mid \bbox{\kappa} \mid t  +\bbox{\kappa}\cdot {\bf
r})] + {\bf A}_2 \exp [\alpha_2 (-\mid \bbox{\kappa} \mid t
+\bbox{\kappa}\cdot {\bf r})] + {\bf A}_3 \exp [\alpha_3
(\bbox{\kappa}\cdot {\bf r})]\quad, \end{equation} with the complex
vectors ${\bf A}_1$, ${\bf A}_2$ and ${\bf A}_3$ and the constants
$\alpha_i$ to be defined from normalization and boundary conditions.  We
have several remarks: a) The plane waves are obtained only if associate
$\lambda = \pm iE$ and $\bbox{\kappa} = \pm i{\bf k}$, what is not
obligatory. It becomes clear that the Maxwell equations may describe
physical states which are different from plane waves, so that the
hypothesis on the quanta of light waves may be regarded as a
particular case only, cf.~[3a,4]; b) The solution with $\lambda =0$ enters
in the general solution of the system of differential equations. It may be
removed only by means of the special choice of boundary conditions; c) In
general, $\bbox{\kappa}$ can be substituted by $-\bbox{\kappa}$ (an analog
of the space inversion transformation in the momentum representation),
{\it i.e.} the solution can be written in several forms, which should be
equivalent in the physical content; d) In the same way one can find the
general solution of the second equation (\ref{2}).

While one can analyze these issues further (and more rigorously) we stop
here in order to be possible to publish an extended
version elsewhere and because of volume restrictions of the journal. But,
below we shall show that {\it non-plane-wave} solutions of the Maxwell's
equations, arise also from different viewpoint~\cite{EVA}, they are {\it
not} zero and that the field related with these
unusual modes   may be {\it irrotational} under certain conditions.
Firstly, we write {\it particular plane-wave} solutions of the Maxwell's
equations in the form\footnote{Here and below the notation may have
nothing to do with the accustomed notation for the vectors of electric and
magnetic fields.}
\begin{equation} {\bf A} ({\bf r}) = \pmatrix{a_1\cr
a_2\cr a_3\cr} e^{i(\omega t  - {\bf k}\cdot {\bf r})} \quad,\quad {\bf B}
({\bf r}) = \pmatrix{b_1\cr b_2\cr b_3\cr} e^{-i(\omega t - {\bf k}\cdot
{\bf r})}\quad, \label{3} \end{equation}
with the objects ${\bf
a}=\mbox{column} (a_1\quad a_2\quad a_3)$ and ${\bf b} = \mbox{column}
(b_1 \quad b_2 \quad b_3)$ at the exponents being the constant {\it
vectors} with respect to the space inversion operation. In order to form
an {\it axial vector} one should add the space-inverted vectors to the
defined ones.\footnote{We still work in the coordinate representation
and want to form an axial vector with respect ${\bf r} \rightarrow -{\bf
r}$. We do not bother the properties of the vectors with respect to ${\bf
k} \rightarrow -{\bf k}$.} Thus, we obtain
\begin{mathletters}
\begin{eqnarray} {\bf C} ({\bf r}) &=& {1\over 2} \left ( \pmatrix{a_1\cr
a_2\cr a_3\cr} e^{i(\omega t - {\bf k}\cdot {\bf r})} -  \pmatrix{a_1\cr
a_2\cr a_3\cr} e^{i(\omega t + {\bf k}\cdot {\bf r})} \right ) =
\pmatrix{a_1\cr a_2\cr a_3\cr} \sin ({\bf k}\cdot {\bf r}) e^{i (\omega t
- \pi/2)} \quad,\quad\\ {\bf D} ({\bf r}) &=& {1\over 2} \left (
\pmatrix{b_1\cr b_2\cr b_3\cr} e^{-i(\omega t - {\bf k}\cdot {\bf r})} -
\pmatrix{b_1\cr b_2\cr b_3\cr} e^{-i(\omega t + {\bf k}\cdot {\bf r})}
\right ) = \pmatrix{b_1\cr b_2\cr b_3\cr} \sin ({\bf k}\cdot {\bf r})
e^{-i (\omega t - \pi/2)} \quad.\quad
\end{eqnarray} \end{mathletters}

We shall further prove the following theorems:

\medskip

{\bf Theorem 1.} {\it The quantity} ${\bf F}={\bf C}\times {\bf D}$ {\it
conserves in time:}
\begin{equation}
{\partial \over \partial t} {\bf F} = 0\quad.
\end{equation}

\smallskip

{\bf Proof.} By the straightforward calculation one can find the explicit
form of the {\it axial vector} ${\bf F}$. Here it is:
\begin{equation}
{\bf F} = \pmatrix{a_2 b_3 - a_3 b_2\cr
a_3 b_1 - a_1 b_3\cr
a_1 b_2 - a_2 b_1\cr} \sin^2 ({\bf k}\cdot {\bf r})
\quad.\label{4}
\end{equation}
By definition the ${\bf a}$ and ${\bf b}$ are the constant {\it vectors}.
Thus, Eq. (\ref{4}) contains no dependence on the time $t$,
so ${\partial {\bf F}\over \partial t} = 0$. Theorem is proven.

\smallskip

{\bf Theorem 2.} {\it If} ${\bf A}$ {\it and} ${\bf B}$ {\it chosen in
the form (\ref{3}) satisfy the Maxwell's equations (\ref{1},\ref{2})
respectively (or vice versa),
the quantity} ${\bf F} = {\bf C} \times {\bf D}$ {\it a) is irrotational;
b) satisfies both equations (\ref{1}) and (\ref{2}); c) is zero in all
space if and only if} ${\bf A}$ {\it or} ${\bf B}$ {\it is zero.}

\smallskip

{\bf Proof.} In order to prove a) and b) it is sufficiently to
prove that $({\bf J} \cdot \bbox{\nabla})^{ij} {\bf F}^j = 0$
because of the operator identity  $\bbox{\nabla} \times \equiv
\mbox{{\bf curl}}$, the definition of the $j=1$ matrices and thanks to the
proven {\bf Theorem 1}.  By direct calculations one comes to
\begin{eqnarray} ({\bf J} \cdot \bbox{\nabla})^{ij} {\bf F}^j &=&
i\bbox{\nabla} \times {\bf F} = i\sin 2({\bf k}\cdot {\bf r}) \,\left \{
{\bf k}\times [ {\bf a} \times {\bf b} ] \right \} \equiv \nonumber\\
&\equiv& i\sin 2({\bf k}\cdot {\bf r}) \left \{ {\bf a} ({\bf k} \cdot
{\bf b}) - {\bf b} ({\bf k}\cdot {\bf a}) \right \} \equiv\nonumber\\
&\equiv& i\sin 2({\bf k}\cdot {\bf r}) \left \{ {\bf a}\times [ {\bf k}
\times {\bf b} ] - {\bf b} \times [ {\bf k} \times {\bf a} ] \right
\}\quad.\label{5} \end{eqnarray} After using the Maxwell's equations
(\ref{1},\ref{2}) one finds ${\bf k} \times {\bf a} = -i\omega {\bf a}$
and ${\bf k} \times {\bf b} = + i\omega {\bf b}$.\footnote{If imply that
${\bf A}$ is a particular solution of (\ref{2}) and ${\bf B}$ is a
particular solution of (\ref{1}) we would have opposite signs in the
written relations.} Substituting these relations to (\ref{5}) we are
convinced that ${\bf F}$ is irrotational and, thus, combining this
statement with the previous one (conservation of ${\bf F}$ in time) we
prove that the quantity ${\bf F}$ satisfies both Maxwell's equations
(\ref{1}) and (\ref{2}).  Following the accustomed terminology it can be
named as ``longitudinal".

Let us now assume that ${\bf F} = {\bf 0}$ in all the space. If ${\bf
a}\neq 0$ and ${\bf b}\neq 0$  this can occur only if ${\bf a}\times
{\bf b}=0$ for the {\it propagating wave} states. By definitions they are
complex {\it vectors}.  So, if denote ${\bf c} = \Re e\, {\bf a}$, ${\bf
d} = \Im m \, {\bf a}$, ${\bf e} = \Re e \,{\bf b}$ and ${\bf f} = \Im m
\, {\bf b}$ we can deduce that in order the searched cross product would
be equal to zero it is necessary
\begin{equation} {\bf c} \times {\bf e} =
+ {\bf d} \times {\bf f} \quad,\quad {\bf d} \times {\bf e} = - {\bf c}
\times {\bf f} \quad, \label{a}
\end{equation}
Let us firstly consider the
case when ${\bf c}$ and ${\bf e}$ are not collinear, ${\bf d}$ and ${\bf
f}$ are not collinear, {\it i.e.} the first relation is not equal to zero.
It can be fulfilled if and only if the {\it real} vectors ${\bf c}$, ${\bf
d}$, ${\bf e}$ and ${\bf f}$ are all coplanar.  Thus, let us choose two
vectors ${\bf c}$ and ${\bf d}$, which are implied to be linear
independent, then other two can be expanded as follows $${\bf e} = a_{11}
{\bf c} + a_{12} {\bf d}\quad,\quad {\bf f} = a_{21} {\bf c} + a_{22} {\bf
d}\quad$$ with real coefficients $a_{ij}$.  Considering ${\bf c}\times
{\bf e}$ and ${\bf d} \times {\bf f}$ we are convinced that the quantity
$a_{12} = - a_{21}$.  Considering the second equation in (\ref{a}) we are
convinced that $a_{11} = a_{22}$.  Thus, ${\bf b}={\bf e}+i{\bf f} =
(a_{11} -ia_{12}) ({\bf c} +i{\bf d})$ and, hence, ${\bf b} \sim c_1
e^{i\beta} {\bf a}$. We have a contradiction with the statement that ${\bf
A}$ and ${\bf B}$, which are {\it not} phase free, satisfy different
Maxwell's equations (\ref{1}) and ({\ref{2}).  Next, if ${\bf d} = \lambda
{\bf c}$ from the set (\ref{a}) we deduce that this can occur if and only
if $\lambda^2 = -1$ what is again in contradiction with the fact that
${\bf c}, {\bf d}, {\bf e}$ and ${\bf f}$ are {\it real} vectors.
Finally, if ${\bf c} = \lambda_1 {\bf e}$ and, then, ${\bf d} = \lambda_2
{\bf f}$ one deduces:  $${\bf d}\times {\bf e} = \lambda_2 {\bf f} \times
{\bf e} = - \lambda_1 {\bf e} \times {\bf f}$$ and, therefore, $\lambda_1
=\lambda_2=\lambda$.  Again, ${\bf b} \sim (1/\lambda) {\bf a}$ and one
has a contradiction with the conditions of the theorem.  So, using the
method of ``from the inverse statement" we can say that ${\bf a}\times
{\bf b}$ cannot be equal to zero and, hence, ${\bf F} \neq {\bf 0}$. The
end of the proof.

\smallskip

{\bf Theorem 3.} {\it If} ${\bf A}$ {\it and} ${\bf B}$ {\it are solutions
of the same equations (\ref{1}) or (\ref{2}) and} $\omega = \pm \mid {\bf
k} \mid$, {\it one can deduce the following} {\bf relation} {\it for the
axial vector} ${\bf F}$ {\it and the corresponding polar} $\mbox{{\bf
curl}}\, {\bf F}$:
\begin{equation} \mbox{{\bf curl}} \, (\mbox{{\bf curl}}
{\bf F}) +4\bbox{\nabla}^2 {\bf F} = 0 \quad.  \label{6}
\end{equation}

\smallskip

{\bf Proof.} The theorem is proving by the direct calculations.
One has
\begin{equation}
\bbox{\nabla} \times {\bf F} =\mp 4i\omega \cot ({\bf k}\cdot {\bf r})
{\bf F} \quad.\label{curl}
\end{equation}
The signs depend on whether ${\bf A}$ and ${\bf B}$ satisfy
simultaneously the first equation (\ref{1}), the sign is ``$-$", or the
second one, the sign is ``$+$".
Next,
\begin{equation}
\bbox{\nabla}^2 {\bf F} = 2{\bf k}^2 \cos 2({\bf k}\cdot {\bf r}) \,
\left [ {\bf a}\times {\bf b} \right ]= 2{\bf k}^2 \, {\cos 2({\bf k}\cdot
{\bf r}) \over \sin^2 ({\bf k}\cdot {\bf r})} {\bf F}\quad,
\end{equation}
and, if one takes into account (\ref{4},\ref{curl}),
\begin{equation} \bbox{\nabla}\times (\bbox{\nabla}\times {\bf F}) =
-8\omega^2 {\cos 2({\bf k}\cdot {\bf r}) \over \sin^2 ({\bf k}\cdot {\bf
r})} {\bf F} \quad.  \end{equation} Substituting these equations in
(\ref{6}) we are convinced in the validity of the theorem.  It is
necessary to stress that  Eq. (\ref{6}) is a {\bf relation}, which was
obtained after taking into account certain constraints between
${\bf k}, {\bf a}, {\bf b}$ and $\omega$. It cannot be
considered as a dynamical {\bf equation}. This is due to the operator
identity $\mbox{{\bf curl}}\, \mbox{{\bf curl}} \equiv \mbox{{\bf grad}}\,
\mbox{{\bf div}} - \bbox{\nabla}^2$. If we rewrite (\ref{6}) with taking
into account this identity we are convinced that the corresponding
equation does {\it not} have solutions unless ${\bf F} = \mbox{const}$,
and/or ${\bf k} \cdot {\bf r} = \pm {\pi \over 4}, \pm
{3\pi \over 4}\ldots$, or ${\bf k} \equiv {\bf 0}$.

\medskip

The conclusion is: the Maxwell's electromagnetic theory looked by a
mathematician/theoretical physicist glance has richer structure comparing
with  views believed since the proposal of quantum field nature of the
light. In the recent series of the papers (see for references~\cite{DVO})
we analyzed its shortcomings and advantages comparing with the more
general Weinberg formalism~\cite{WEIN}.  The question, whether the former
is equivalent to the latter, is  still required further rigorous
elaboration.

{\it Acknowledgments.}
We acknowledge discussions with Profs. D. V. Ahluwalia, A. E. Chubykalo,
and A. F. Pashkov. We are indebted to Prof. H. M\'unera
for his very helpful and important preprints sent to us before
publications.  This essay is submitted to the special issue of ``Apeiron"
devoted to the longitudinal solutions of relativistic equations.
We are thankful to C. Roy Keys, the Editor-in-Chief  for kind invitation
to write it and for arranging this volume.

York University, Canada and the Indian Statistical Institute are
thanked for the award of visiting professorships to M.W.E.
Other author (V.V.D.) is grateful to Zacatecas University for a
professor position.

\end{document}